\documentclass[1p,number,sort&compress]{elsarticle}
\usepackage{graphicx}
\usepackage{amsmath}
\usepackage{subfigure}
\usepackage{amsfonts}

\journal{Physica A}
\begin{document}

\begin{frontmatter}
\title{Binary collision approximation for multi-decorated granular chains}
\author[ufpb]{Eduardo Andr\'e de F. Bragan\c ca} 
\author[ufpb]{Alexandre Rosas}
\ead{arosas@fisica.ufpb.br}
\author[ucsd]{Katja Lindenberg}
\address[ufpb]{Departamento de F\'{\i}sica, CCEN, Universidade Federal da Para\'{\i}ba, Caixa Postal 5008, 58059-900, Jo\~ao Pessoa, Brazil.}
\address[ucsd]{Department of Chemistry and Biochemistry and BioCircuits Institute, University of California San Diego, La Jolla, California 92093-0340, USA.
}

%\date{\today}

\begin{abstract}
We study pulse propagation along decorated tapered granular chains without precompression. Our goal is to generalize the results obtained in our previous work, by analyzing a decorated chain with  an arbitrary number of small grains between the large ones. Making use of an effective description, where the original decorated tapered chain is replaced by a non-decorated tapered chain with effective masses interacting via an effective potential, and applying the binary collisions approximation, we calculate the residence time of the pulse on each effective large grain. We also present the comparison between the numerical integration of the equations of motion and our analytical predictions which show the agreement to be very good for the pulse velocity, albeit only qualitatively for the velocity of the grains.
\end{abstract}
\end{frontmatter}
%\pacs{46.40.Cd,43.25.+y,45.70.-n}

%\maketitle

\section{Introduction}
Pulse propagation in granular chains has attracted the attention of many scientists since it was first shown that an initial impulse imparted at one end of the chain could give rise to solitary waves \cite{nesterenko,nesterenko-1}. Different types of chains have been studied since then. In particular, polydispersity is one of the characteristics of granular chains which has been considered in regular fashion, as in tapered ~\cite{nakagawa,wu02,robertPRE05,sokolowAPL05,senPHYA01,meloPRE06,jobGM07,robertPRL06,Harbola,Harbola-2}, and/or randomly structured chains~\cite{nesterenko,nesterenkobook,SokolowSen07,ChenWang07,Harbola-3}.

Recently~\cite{Harbola-2} we studied pulse propagation in a mono-decorated tapered chain, that is, a chain of large grains with radii that decrease systematically, with a small grain between each pair of large grains.
In this contribution, we are interested in pulse propagation in a decorated granular chain decorated with an arbitrary number of small grains between the large ones. In particular, we are interested in the velocity profiles and the residence time of the pulse as the initial disturbance travels along the chain.
Our main purpose is to apply the binary collision approximation to describe pulse propagation along the chain. This theory has been successfully employed to analyze the pulse propagation through several types of chains, including monodisperse \cite{Rosas}, tapered \cite{Harbola}, o-rings with and without precompression \cite{Italo}, decorated \cite{Harbola-2} and randomly decorated \cite{Harbola-3}, and Y-shaped chains \cite{Daraio,Chen}. However, as has already been pointed out in \cite{Harbola-2}, the binary collision approximation cannot be directly applied to decorated chains because the energy and momentum transfer occur through several oscillations of the small grains instead of a single forward transfer, as assumed by the approximation. Therefore we proceed to obtain an effective chain composed of only large grains. These new large grains have effective masses and interact via effective potentials which we calculate here.

This paper is organized as follows. We describe the interaction model in Section~\ref{sec:model}. In Section~\ref{sec:effectivedynamics} we obtain the effective potential, while in Section~\ref{sec:binarycollision} we present the binary collision approximation for the effective chain. In both sections we show comparisons of the theoretical predictions with the numerical integration of the equations of motion. Finally, we present concluding remarks in Section~\ref{sec:conclusion}.

\section{Model}
\label{sec:model}
We consider chains of grains placed along a line so that they just touch their neighbors (there is no precompression), and all but the leftmost particle are at rest. The equation of motion for the $k$-th grain (except for the first and last grains) is
\begin{equation}
  \begin{split}
 M_k\frac{d^2y_k}{d{\tau}^2}&= {a}{r}_{k-1}^\prime(y_{k-1}-y_{k})^{n-1}\theta(y_{k-1}-y_k)
                            \\&-{a}{r}_{k}^\prime(y_k-y_{k+1})^{n-1}\theta(y_k-y_{k+1}).
\label{eq:motion}
\end{split}
\end{equation}
For the first (last) grain, the first (second) term in the right-hand absent. $M_{k}$ is the mass of grain $k$, $y_{k}$ is its displacement from its initial position and $a$ is a constant determined by Young's modulus and Poisson's ratio \cite{Landau,Hertz}. The Heaviside function $\theta(y)$ ensures that the elastic interaction between grains only exists if they are in contact.  The constant $r'_k$ is given by
\begin{equation}
  r_k^\prime = \left(\frac{2 R_k^\prime R_{k+1}^\prime}{R_k^\prime +R_{k+1}^\prime }\right)^{1/2},
\end{equation}
where $R^\prime_{k}$ is the principal radius of curvature of the surface of grain $k$ at the point of contact with grain $k+1$. The initial velocity $V_{1}$ of the leftmost particle ($k=1$) provides the initial perturbation which will give rise to a propagating pulse~\cite{nesterenkobook}. For convenience, we introduce the dimensionless quantity
\begin{eqnarray}
\label{eq:rescaled}
 \alpha \equiv \left[\frac{M_{1}{V_{1}^{2}}}{a(R^{\prime}_{1})^{n + 1/2}}\right]^{1/n},
\end{eqnarray}
and the rescaled quantities $x_{k}$, $t$, $m_{k}$, and $R_{k}$ via the relations
\begin{equation}
\label{eq:relations}
y_{k} = R^\prime_{1} \alpha^{1/n}x_{k}, \quad \tau =\frac{R^{\prime}_{1}}{V_{1}}\alpha^{1/n}t, \quad 
R^{\prime}_{k} = R^{\prime}_{1}R_{k}, \quad M_{k} = M_{1}m_{k},
\end{equation}
so that the Eq. (\ref{eq:motion}) can be rewritten as
\begin{equation}
\label{eq:motionrescaled}
m_{k} \ddot{x}_{k}=r_{k-1}(x_{k-1} - x_{k})^{n-1}\theta(x_{k-1} - x_{k})
                  -r_{k}(x_{k} - x_{k+1})^{n-1}\theta(x_{k} - x_{k+1}).
\end{equation}
A dot denotes a derivative with respect to $t$, and
\begin{equation}
\label{eq:rreescaled}
r_{k} = \left( \frac{2R_{k}R_{k+1}}{R_{k} + R_{k+1}} \right)^{1/2}.
\end{equation}

In the rescaled variables, the initial velocity is unity, i.e., $v_{1}(t=0)=1$, while the velocity of the $k$-th grain in the unscaled variables is simply $V_{1}$ times its velocity in the scaled variables.

\section{Effective dynamics}
 \label{sec:effectivedynamics}
\begin{figure}[!ht]
\centering
\scalebox{0.4}{\includegraphics{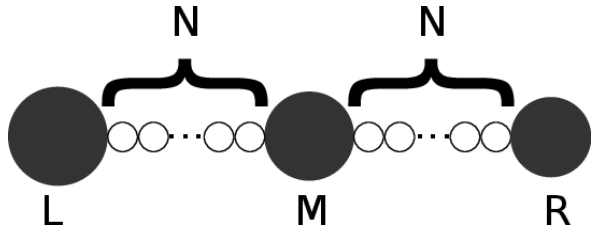}}
\caption{Illustration of a short tapered chain composed of three large grains of decreasing radii and $N$ small grains placed between each pair of large grains.} \label{fig:shortchain}
\end{figure}

When an initial impulse is imparted at one end of a tapered chain decorated with $N$ small grains between each pair of large grains, energy and momentum travel along the chain. A large grain pushes the subsequent small grains which oscillate and transfer momentum and energy. Despite the fact that the dynamics of these oscillations is not easily described, it is possible to approximately describe the large grain dynamics by designing an effective chain composed of large grains with effective masses, interacting via effective potentials. In order to design the effective chain, we follow a procedure similar to that in~\cite{Harbola-2} and write the position of the small grains as
\begin{eqnarray}
  \label{eq:approx}
x_{k}(t)=\bar{x}_{k}(t)+g_{k}(t),
\end{eqnarray}
where $\bar{x}_{k}(t)$ represents the equilibrium position of the small grains at any time, and $g_{k}(t)$ is its deviation from $\bar{x}_{k}(t)$. Our main approximation is to assume that $g_{k}$ is small enough to be neglected. The dynamics of a short chain composed of three large grains and $N$ small grains placed between each pair of large grains (see Fig.~\ref{fig:shortchain}), is governed by the following set of equations of motion
\begin{subequations}
\label{eq:motionNsmallgrains}
\begin{eqnarray}
\label{eq:motionxL}
M_{L} \ddot{x}_{L} &=& -\mathcal{R}_{L}(x_{L}-\bar{x}_{1})^{n-1},\\
0 &=& \mathcal{R}_{L}(x_{L}-\bar{x}_{1})^{n-1}-\xi(\bar{x}_{1}-\bar{x}_{2})^{n-1},\\
0 &=& \xi(\bar{x}_{1}-\bar{x}_{2})^{n-1}-\xi(\bar{x}_{2}-\bar{x}_{3})^{n-1},\\
  & \vdots & \nonumber \\
0 &=& \xi(\bar{x}_{N-2}-\bar{x}_{N-1})^{n-1}-\xi(\bar{x}_{N-1}-\bar{x}_{N})^{n-1},\\
0 &=& \xi(\bar{x}_{N-1}-\bar{x}_{N})^{n-1}-\mathcal{R}_M(\bar{x}_{N}-x_{M})^{n-1},\\
\label{eq:motionxM}
M_{M}\ddot{x}_{M} &=& \mathcal{R}_{M}(\bar{x}_{N}-x_{M})^{n-1}-\mathcal{R}_{M}(x_{M}-\bar{x}_{N+1})^{n-1},\\
                0 &=& \mathcal{R}_{M}(x_{M}-\bar{x}_{N+1})^{n-1}-\xi(\bar{x}_{N+1}-\bar{x}_{N+2})^{n-1},\\
                0 &=& \xi(\bar{x}_{N+1}-\bar{x}_{N+2})^{n-1}-\xi(\bar{x}_{N+2}-\bar{x}_{N+3})^{n-1},\\
&\vdots&\nonumber \\
0 &=& \xi(\bar{x}_{2N-2}-\bar{x}_{2N-1})^{n-1}-\xi(\bar{x}_{2N-1}-\bar{x}_{2N})^{n-1},\\
0 &=& \xi(\bar{x}_{2N-1}-\bar{x}_{2N})^{n-1}-\mathcal{R}_{R}(\bar{x}_{2N}-x_{R})^{n-1},\\
\label{eq:motionxR}
M_{R}\ddot{x}_{R} &=& \mathcal{R}_{R}(\bar{x}_{2N}-x_{R})^{n-1},
\end{eqnarray}
\end{subequations}
where
\begin{equation}
\xi=r^{1/2} \label{eq:xi} \quad \mathrm{and}	 \quad \mathcal{R}_{\alpha }=\left(\frac{2R_{\alpha }r}{R_{\alpha }+r}\right)^{1/2}
\end{equation}
for $ \alpha = L, M$ and $R$, are obtained from the Eq. (\ref{eq:rreescaled}). We have omitted the Heaviside function for the ease of presentation but they are, of course, implicit.
Therefore, the equations for the small grains become
\begin{subequations}
\label{eq:averagemotion}
\begin{eqnarray}
\label{eq:x1bar}
\bar{x}_{1}&=&\left(\frac{\gamma_{L}}{1+\gamma_{L}}\right)x_{L} + \left(\frac{1}{1+\gamma_{L}}\right)\bar{x}_{2}, \\ 
\bar{x}_{2}&=&\frac{1}{2}(\bar{x}_{1}+\bar{x}_{3}),\\
&\vdots & \nonumber \\
\bar{x}_{N-1}&=&\frac{1}{2}(\bar{x}_{N-2}+\bar{x}_{N}),\\
\label{eq:xNbar}
\bar{x}_{N}&=&\left(\frac{1}{1+\gamma_{M}}\right)\bar{x}_{N-1}+\left(\frac{\gamma_{M}}{1+\gamma_{M}}\right)x_{M}, \\ 
\label{eq:xN1bar}
\bar{x}_{N+1}&=&\left(\frac{\gamma_{M}}{1+\gamma_{M}}\right)x_{M} + \left(\frac{1}{1+\gamma_{M}}\right)\bar{x}_{N+2}, \\ 
\bar{x}_{N+2}&=&\frac{1}{2}(\bar{x}_{N+1}+\bar{x}_{N+3}),\\
&\vdots & \nonumber \\
\bar{x}_{2N-1}&=&\frac{1}{2}(\bar{x}_{2N-2}+\bar{x}_{2N}),\\
\label{eq:x2Nbar}
\bar{x}_{2N}&=&\left(\frac{1}{1+\gamma_{R}}\right)\bar{x}_{2N-1}+\left(\frac{\gamma_{R}}{1+\gamma_{R}}\right)x_{R}, 
\end{eqnarray}
\end{subequations}
where
\begin{equation}
\gamma_{L}=\left(\frac{\mathcal{R}_{L}}{\xi}\right)^{1/(n-1)}, 	\quad \gamma_{M}=\left(\frac{\mathcal{R}_{M}}{\xi}\right)^{1/(n-1)} \quad 
\mathrm{and} \quad \gamma_{R}=\left(\frac{\mathcal{R}_{R}}{\xi}\right)^{1/(n-1)}.
\label{eq:gamma}
\end{equation}

Note that the equilibrium displacements of the small grains in contact only with other small grains is just the average of the displacements of their neighbours, since they have the same size and mass. Hence, these grains obey the equation
\begin{equation}
  \bar{x}_{k+1} = \frac{1}{2} \left ( \bar{x}_k + \bar{x}_{k+2} \right ).
  \label{eq:average}
\end{equation}
At this point we are left with a linear system of 2N unknowns $(\bar{x}_{1}, ..., \bar{x}_{N}, \bar{x}_{N+1}, ..., \bar{x}_{2N}),$ and 2N equations. To verify that
\begin{equation}
\label{eq:solution1}
\bar{x}_{p}=\frac{\gamma_{L}x_{L}+(N-p)\gamma_{L}\gamma_{M} x_{L}+(p-1)\gamma_{L}\gamma_{M}x_{M} 
+\gamma_{M}x_{M}}{\gamma_{L}+\gamma_{M}+ (N-1)\gamma_{L}\gamma_{M}},
\end{equation}
and
\begin{equation}
 \label{eq:solution2}
 \bar{x}_{N+p}=\frac{\gamma_{M}x_{M}+(N-p)\gamma_{M}\gamma_{R} x_{M}+(p-1)\gamma_{M}\gamma_{R}x_{R} 
+\gamma_{R}x_{R}}{\gamma_{M}+\gamma_{R}+ (N-1)\gamma_{M}\gamma_{R}},
\end{equation}
for $p=1,2, ..., N$, are solutions of Eq. (\ref{eq:averagemotion}), it is only necessary to substitute these values into Eqs. (\ref{eq:x1bar}), (\ref{eq:xNbar}), (\ref{eq:xN1bar}) and (\ref{eq:x2Nbar}) with the appropriate value of $p$ and $\bar{x}_{k}, \bar{x}_{k+1}$ and $\bar{x}_{k+2}$ in the general Eq. (\ref{eq:average}).

Furthermore, note that the sum of all the equations of motion gives
\begin{equation}
\label{eq:sumeqmotion}
M_{L}\ddot{x}_{L} + m\sum_{p=1}^{N} \ddot{\bar{x}}_{p}+  M_{M}\ddot{x}_{M} +m\sum_{p=1}^{N} \ddot{\bar{x}}_{N+p}
+M_{R}\ddot{x}_{R}=0
\end{equation}
as a consequence of the conservation of momentum. On the other hand, summing up Eqs. (\ref{eq:solution1}) and (\ref{eq:solution2}), we have respectively
\begin{eqnarray}
\label{eq:sump}
\sum_{p=1}^{N} \bar{x}_{p}&=&\dfrac{N\left[1+\dfrac{1}{2}(N-1)\gamma_{M}\right]\gamma_{L}x_{L}+N\left[1+\dfrac{1}{2}(N-1)\gamma_{L}\right]\gamma_{M}x_{M}}{\gamma_{L}+\gamma_{M}+(N-1)\gamma_{L}\gamma_{M}},\\
\label{eq:sumq}
\sum_{p=1}^{N} \bar{x}_{N+p}&=&\dfrac{N\left[1+\dfrac{1}{2}(N-1)\gamma_{R}\right]\gamma_{M}x_{M}+N\left[1+\dfrac{1}{2}(N-1)\gamma_{M}\right]\gamma_{R}x_{R}}{\gamma_{M}+\gamma_{R}+(N-1)\gamma_{M}\gamma_{R}}.
\end{eqnarray}
Hence, taking the second derivative of Eqs. (\ref{eq:sump}) and (\ref{eq:sumq}) we may rewrite Eq. (\ref{eq:sumeqmotion}) as
\begin{equation}
\mu_L \ddot{x}_{L}+\mu_M \ddot{x}_{M}+\mu_R \ddot{x}_{R}=0,
\end{equation}
where we have defined the effective masses
\begin{subequations}
\begin{align}
 \label{eq:massleft}
 \mu_{L}&= M_{L}+\frac{mN\left[1+\frac{1}{2}(N-1)\gamma_{M}\right]\gamma_{L}}{\gamma_{L}+\gamma_{M}+(N-1)\gamma_{L}\gamma_{M}},\\
 \label{eq:massmiddle}
 \mu_{M}&=\frac{mN\left[1+\frac{1}{2}(N-1)\gamma_{L}\right]\gamma_{M}}{\gamma_{L}+\gamma_{M}+(N-1)\gamma_{L}\gamma_{M}}+M_{M} 
+\frac{mN\left[1+\frac{1}{2}(N-1)\gamma_{R}\right]\gamma_{M}}{\gamma_{M}+\gamma_{R}+(N-1)\gamma_{M}\gamma_{R}},\\
 \label{eq:massright}
 \mu_{R}&=\frac{mN\left[1+\frac{1}{2}(N-1)\gamma_{M}\right]\gamma_{R}}{\gamma_{M}+\gamma_{R}+(N-1)\gamma_{M}\gamma_{R}}+M_{R}.
\end{align}
\label{eq:effectivemass}
\end{subequations}
Thus, the dynamics of a decorated chain of three large grains with N small grains between each pair can be approximately described by an effective chain of three large grains labeled ``left'' (L), ``middle'' (M), and ``right'' (R). It is easy to see that these equations reduce to Eq. (23) of~\cite{Harbola-2}

Hence we are left with the equations of motion (\ref{eq:motionxL}), (\ref{eq:motionxM}) and (\ref{eq:motionxR}). In order, to take into account the effect of the small grains, we use the effective masses instead of the masses of the original large grains. In addition, we use Eqs. (\ref{eq:solution1}) and (\ref{eq:solution2}) to replace $\bar{x}_{1}, \; \bar{x}_{N}, \; \bar{x}_{N+1}$ and $\bar{x}_{2N}$, so that (we also make use of Eq. (\ref{eq:gamma}) to write the following equations)
\begin{subequations}
\begin{eqnarray}
 \label{eq:leftmass}
\mu_{L}\ddot{x}_{L}&=&-\frac{\left(\gamma_{L}^{n-1}\gamma_{M}^{n-1}\xi\right)(x_{L}-x_{M})^{n-1}}{\left[\gamma_{L}+\gamma_{M}+(N-1)\gamma_{L}\gamma_{M}\right]^{n-1}},\\
 \label{eq:middlemass}
\mu_{M}\ddot{x}_{M}&=&\frac{\left(\gamma_{L}^{n-1}\gamma_{M}^{n-1}\xi\right ) (x_{L}-x_{M})^{n-1}}{\left[\gamma_{L}+\gamma_{M}+(N-1)\gamma_{L}\gamma_{M}\right]^{n-1}}
-\frac{\left(\gamma_{M}^{n-1}\gamma_{R}^{n-1}\xi\right)(x_{M}-x_{R})^{n-1}}{\left[\gamma_{M}+\gamma_{R}+(N-1)\gamma_{M}\gamma_{R}\right]^{n-1}},\\
\label{eq:rightmass}
\mu_{R}\ddot{x}_{R}&=&\frac{\left(\gamma_{M}^{n-1}\gamma_{R}^{n-1}\xi\right)(x_{M}-x_{R})^{n-1}}{\left[\gamma_{M}+\gamma_{R}+(N-1)\gamma_{M}\gamma_{R}\right]^{n-1}}.
\end{eqnarray}
\end{subequations}

These equations define an effective two-particle potential given by
\begin{equation}
  \label{eq:potencial}
V_{eff}=\frac{\zeta_{k}(n)}{n}(x_{k}-x_{k+1})^{n},
\end{equation}
where
\begin{equation}
 \label{eq:zeta}
\zeta_{k}(n)=\frac{\gamma_{k}^{n-1}\gamma_{k+1}^{n-1}\xi}{\left[\gamma_{k}+\gamma_{k+1}+(N-1)\gamma_{k}\gamma_{k+1}\right]^{n-1}},
\end{equation}
and the three grains in the effective chain are relabeled as $(L,M,R)\rightarrow(k-1,k,k+1)$. For the particular case $N=1$, the previous equation reduces to
\begin{equation}
  \zeta_{k}(n)=\frac{1}{\frac{1}{\gamma_{k}\xi}+\frac{1}{\gamma_{k+1}\xi}},
\end{equation}
which is Eq. (25) of~\cite{Harbola-2}.

\begin{figure*}[!ht]
\vspace{.6cm}
\hspace*{-1.8cm}
$\begin{array}{cc}
\rotatebox{0}{\scalebox{.6}{\includegraphics{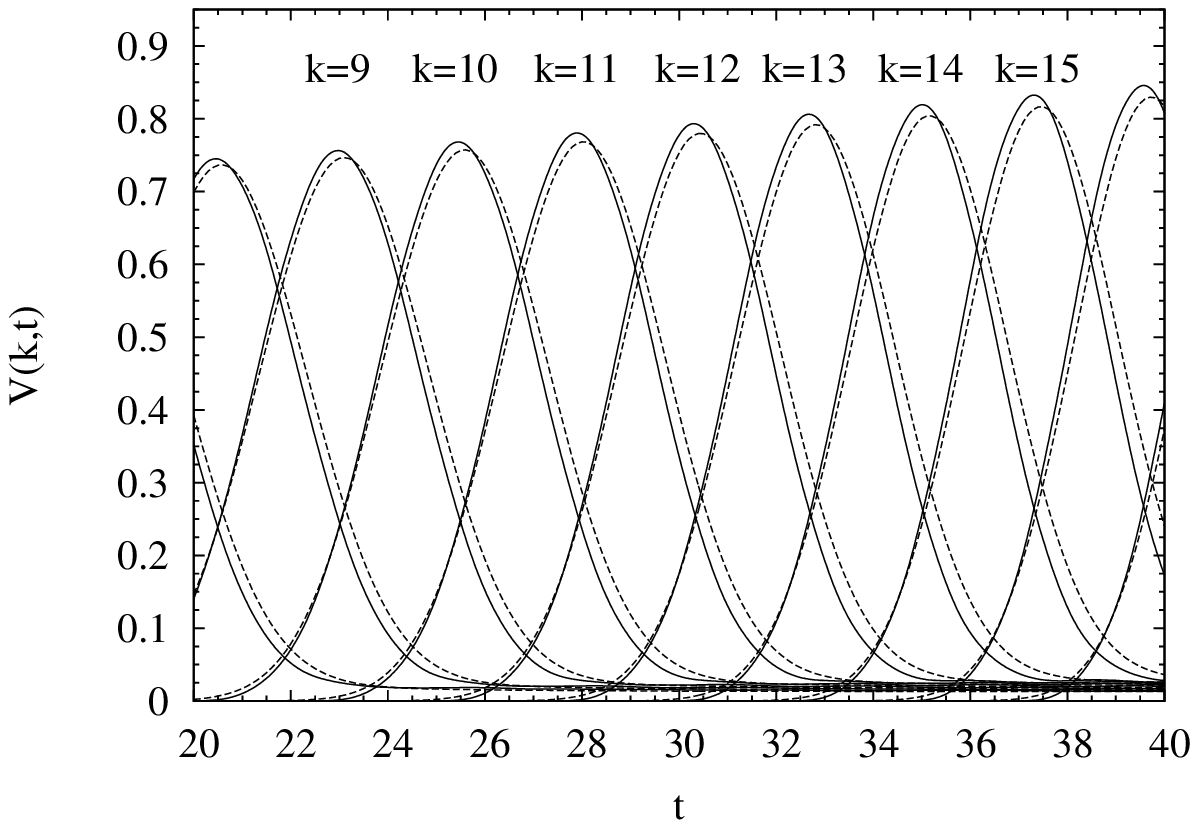}}}&
\rotatebox{0}{\scalebox{.6}{\includegraphics{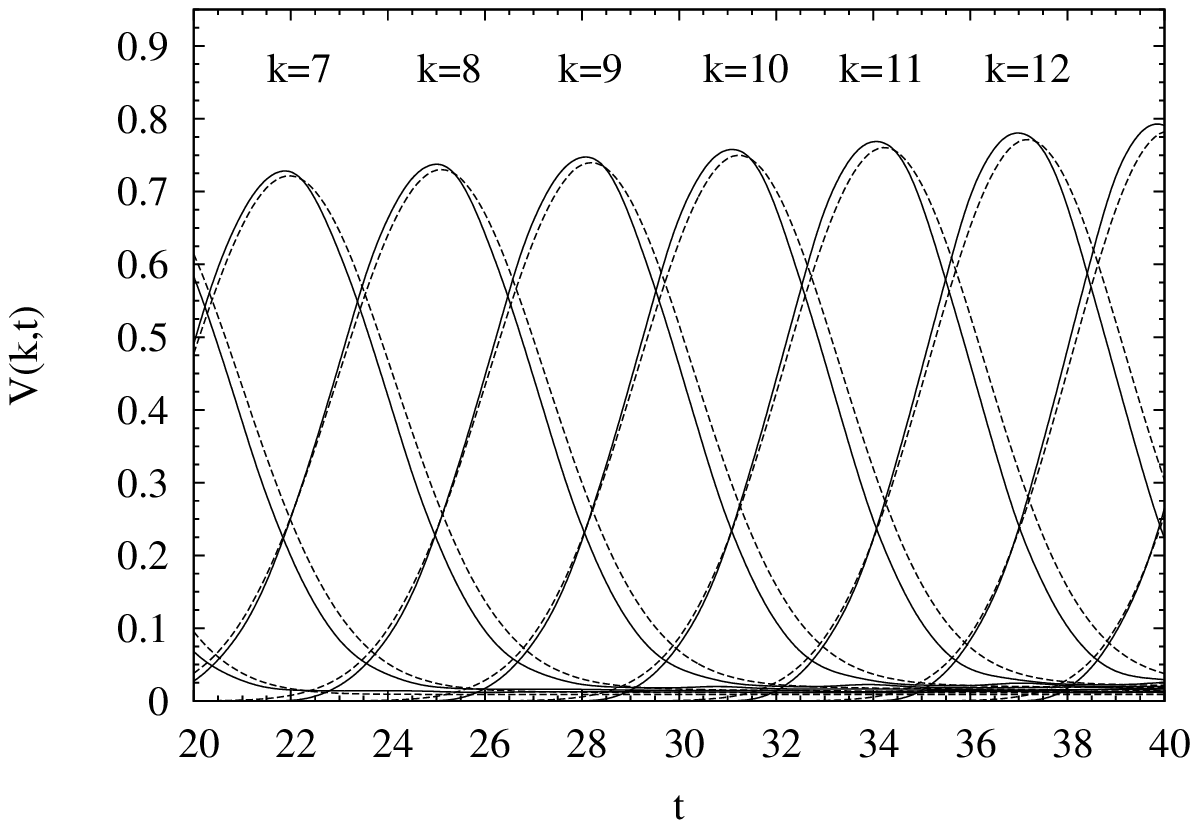}}}\\
\rotatebox{0}{\scalebox{.6}{\includegraphics{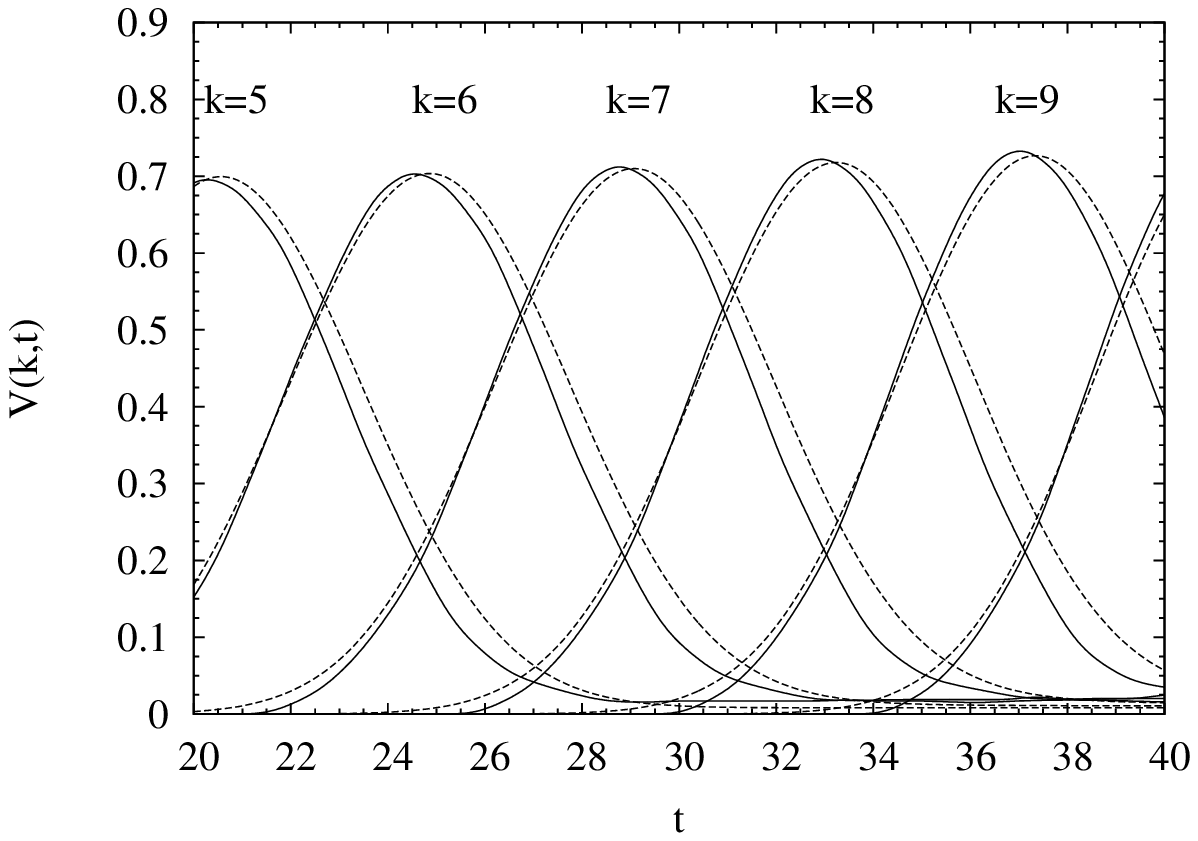}}}&
\rotatebox{0}{\scalebox{.6}{\includegraphics{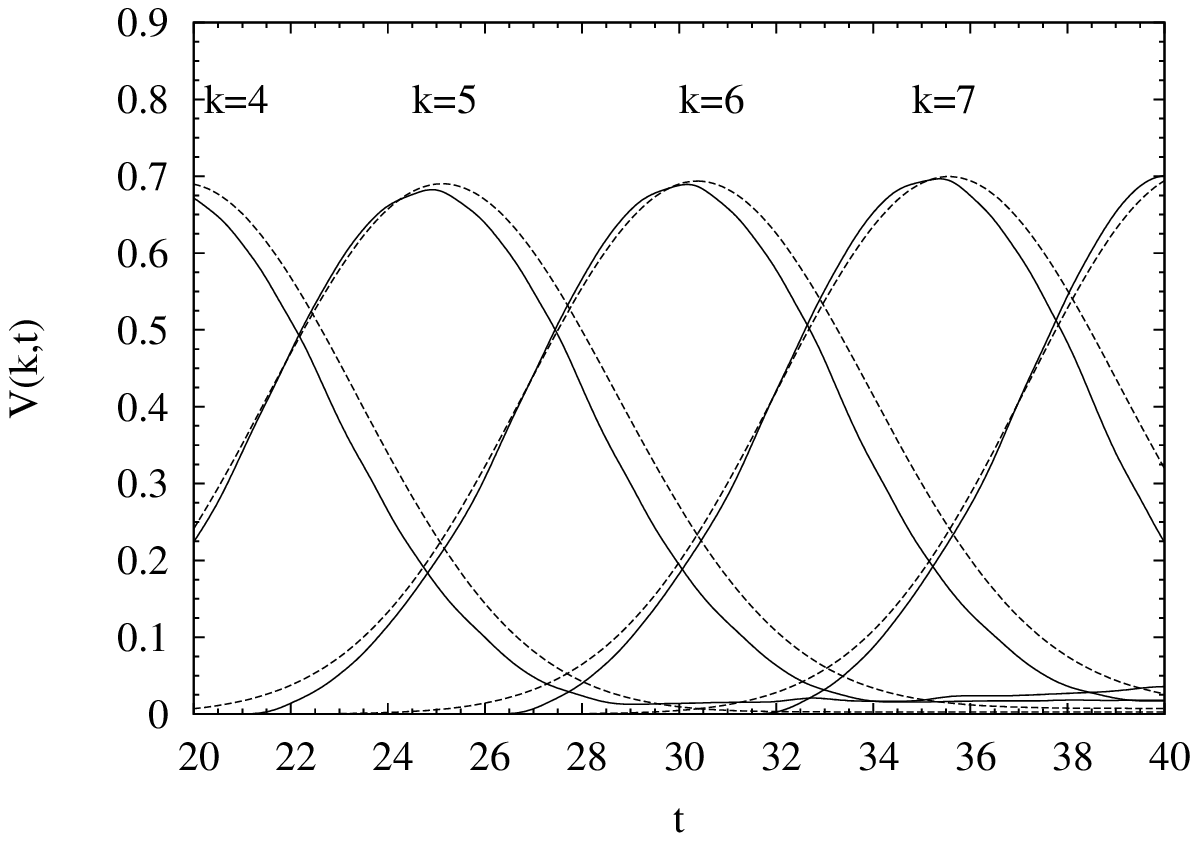}}}
\end{array}$
\caption{Evolution of velocities of the grains in the decorated (solid curves) and effective (dashed curves) chains. Here, the radii of the small grains between large ones is $r=0.3,$ the number of small grains is $N=2$ (top left), $3$ (top right), $5$ (bottom left) and $7$ (bottom right), and the tapering parameter is $S=0.01$.}
 \label{fig:velocitylarge}
\end{figure*}

In summary, the decorated chain of large grains with $N$ small grains between them can be approximately described by an effective chain of only large grains with effective masses and interacting via the potential of Eq. (\ref{eq:potencial}). Note that the effective mass $\mu_{M}$ is modified by two groups of small grains, one on each side, so that it describes all particles not at the edges of the chain, while the effective masses $\mu_{L}$ and $\mu_{R}$ describe the leftmost and rightmost particles, respectively, which are modified by only one group of small grains.  In Fig. \ref{fig:velocitylarge}, the time evolution of the original equations of motion and that of the effective chain are compared. In this figure, we show the velocities of a few large grains which present a few remarkable features. First, we can see that the effective description captures the behavior of the large grains but the difference between the original and the effective chains increase with the number of decorating particles. This is not surprising since we are approximating more and more oscillatory terms as we increase the number of small grains. It is also worth noticing that as more small grains are introduced in the chain, the pulse becomes slower and broader in time.

\section{Binary collision approximation}
\label{sec:binarycollision}
We now use the binary collision approximation to study the propagation of the pulse along a decorated chain via the effective description of Sec. \ref{sec:effectivedynamics}. 
The binary collision approximation supposes that the pulse propagation in the chain occurs via a succession of two-particle collisions. First the leftmost particle $k=1$, with velocity $v_{1}=1$ in the scaled variables, collides with the particle $k=2$, which is initially at rest. This latter particle then acquires a velocity $v_{2}$ and collides with the next particle, $k=3$, also initially at rest, and so on. Using energy and momentum conservation, and remembering that the grains in the chain have effective mass $\mu_{k}$, it can be shown that after the collision of the grains $k-1$ and $k$, the velocity of grain $k$ is given by~\cite{Harbola}
\begin{eqnarray}
 \label{eq:Vbinary}
v_{k} = \prod\limits_{k^{'}=1}^{k-1} \frac{2}{1+\frac{\mu_{k^{'}+1}}{\mu_{k^{'}}}}.
\end{eqnarray}

In order to calculate the residence time of the pulse, that is, the time required for grain $k+1$ to become faster than grain $k$ after the start of the collision, we write the two-particles equations of motion as
\begin{eqnarray}
 \label{eq:difference}
\ddot{x}_{k}=-\frac{\zeta_{k}(n)}{\mu_{k}}(x_{k}-x_{k+1})^{n-1}, \nonumber \\
\ddot{x}_{k+1}=\frac{\zeta_{k}(n)}{\mu_{k+1}}(x_{k}-x_{k+1})^{n-1},
\end{eqnarray}
where $\zeta_{k}$ is defined in Eq. (\ref{eq:zeta}). Next, we introduce the relative variable
\begin{eqnarray}
 \label{eq:z}
z_{k}=x_{k}-x_{k+1},
\end{eqnarray}
and subtract the two equations of motion obtaining the equation for the evolution of the relative variable
\begin{equation}
 \label{eq:zmotion}
\ddot{z}_{k}=-\frac{\zeta_{k}(n)}{\mathcal{M}_{k}}z_{k}^{n-1}.
\end{equation}
Here
\begin{eqnarray}
 \label{eq:reducedmass}
\mathcal{M}_{k}=\frac{\mu_{k}\mu_{k+1}}{\mu_{k}+\mu_{k+1}}
\end{eqnarray}
is the reduced mass of the two-particle system.

Equation (\ref{eq:zmotion}) is the equation of motion of a particle with effective mass $\mathcal{M}_{k}$ subjected to the potential $\frac{\zeta_{k}(n)}{n}z^{n}$. The initial conditions are $z_{k}(0)=0,$ since there is no precompression, and $\dot{z}_{k}=v_{k}$, since the velocity of grain $k+1$ is zero before the collision. The energy conservation condition then reads
\begin{equation}
  \label{eq:energy}
\frac{1}{2}\dot{z}^{2}_{k}(t)+\frac{\zeta_{k}(n)}{n\mathcal{M}_{k}}z^{n}_{k}(t)=\frac{1}{2}\dot{z}_{k}^{2}(0).
\end{equation}

Now we can calculate the residence time as
\begin{equation}
T_{k}=\int^{z_{k}^{max}}_{0}\frac{dz_{k}}{\dot{z}_{k}} = \int_{0}^{z_{k}^{max}} \frac{dz_{k}}{\left(\dot{z}_{k}^{2}(0)-\frac{2\zeta_{k}(n)}{n\mathcal{M}_{k}}z_{k}^{n}\right)^{1/2}},
\end{equation}
where $z_{k}^{max}$ is the maximum compression, that is, the compression for which the velocities of particles $k$ and $k+1$ are the same, and, consequently the relative velocity is zero. The maximum compression can readily be obtained by imposing that $\dot{z}_k(t) = 0$ in Eq. (\ref{eq:energy}). The result is 
\begin{eqnarray}
z_{k}^{max}=\left(\frac{n\mathcal{M}_{k}}{2\zeta_{k}(n)}\dot{z}_{k}^{2}(0)\right)^{1/n}.
\end{eqnarray}

The integral can be performed exactly, yielding
\begin{eqnarray}
 \label{eq:residencetime}
T_{k}=\sqrt{\pi}\left(\frac{n\mathcal{M}_{k}}{2\zeta_{k}(n)}\right)^{1/n}[\dot{z}_{k}(0)]^{-1+2/n} \frac{\Gamma(1 + 1/n)}{\Gamma(1/2 + 1/n)}.
\end{eqnarray}

This is the main result of the binary collision approximation; it will be compared with the results of the numerical integration of the equations of motion of the original chain. Equation~(\ref{eq:residencetime}) is valid for any kind of tapering of the large grains (and also for monodispersed chains) and any value $n>2.$ However, we focus our numerical comparisons on spherical grains, i.e., $n=5/2,$ since, experimentally, this is the most studied case. In order to apply Eq. (\ref{eq:residencetime}) to a particular case, we need the specific tapering rule, which defines the effective masses $\mu_k$ [Eq. (\ref{eq:effectivemass})], as well as the effective potential constant [Eq. (\ref{eq:zeta})], and the velocity of the impacting grain. In the spirit of the binary collision approximation, this is the velocity of the incoming grain at the end of the previous collision, given by Eq. (\ref{eq:Vbinary}). Here, we consider a forward linearly tapered chain (60 grains long) decorated with $N$ small grains of radii $r$. That is, the radius of the $k$-th large grain is $R_{k}=1-S(k-1),$ where $S$ is the tapering parameter \cite{Harbola}. 

\begin{figure}[!ht]
\centering
\rotatebox{0}{\scalebox{.7}{\includegraphics{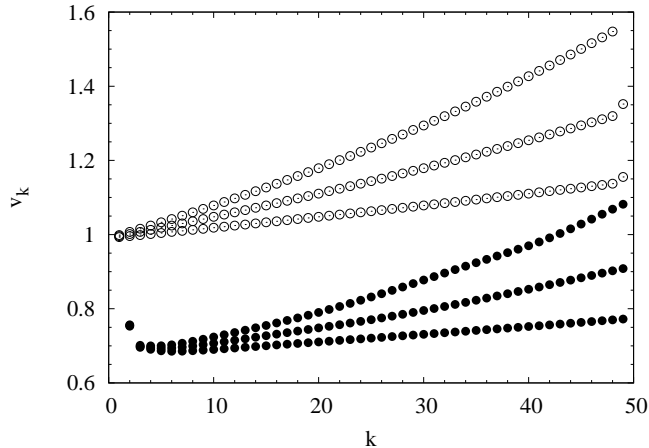}}}
\caption{Change in the velocity pulse amplitude as a function of $k$. The filled and empty circles represent the numerical and binary collision approximation results, respectively. Here $N=3$ and $S=0.002, 0.004, 0.006$ from bottom to top. In all cases $r=0.3$.} \label{fig:amplitude}
\end{figure}

In Fig. \ref{fig:amplitude} we show the velocity pulse amplitude as a function of large grain number. Numerically, the pulse amplitude at grain $k$ is determined as its maximum velocity. Theoretically, we use Eqs. (\ref{eq:Vbinary}) and (\ref{eq:effectivemass}), recursively, remembering that, for spheres, the masses of the large grains are $M_k = 4 \pi R_k^3/3$ and the masses of the small grains are $m = 4 \pi r^3/3$. The failure of the binary collision approximation in obtaining the correct values of the pulse is obvious, but it correctly captures the trend of the pulse amplitude. This is a well known problem of the binary collision approximation, which may be tackled using a numerical cum analytical approach, as proposed in \cite{machado}. Another important characteristic of this figure is the increasing pulse amplitude along the chain due to the decrease of the masses of the large grains in the tapered chain.
%%%%%%%%%%%%%%%%%%%%%%%%%%%%%%%%%%%%%%%%%%%%%%%%%%%%%%%%%%%%%%%%%%%%%%%%
%%%%%%%%%%%%%%%%%%%%%%%%%%%%%%%%%%%%%%%%%%%%%%%%%%%%%%%%%%%%%%%%%%%%%%%%%%

\begin{figure*}[!ht]
\vspace{.6cm}
\hspace*{-1.8cm}
$\begin{array}{cc}
\rotatebox{-90}{\scalebox{.30}{\includegraphics{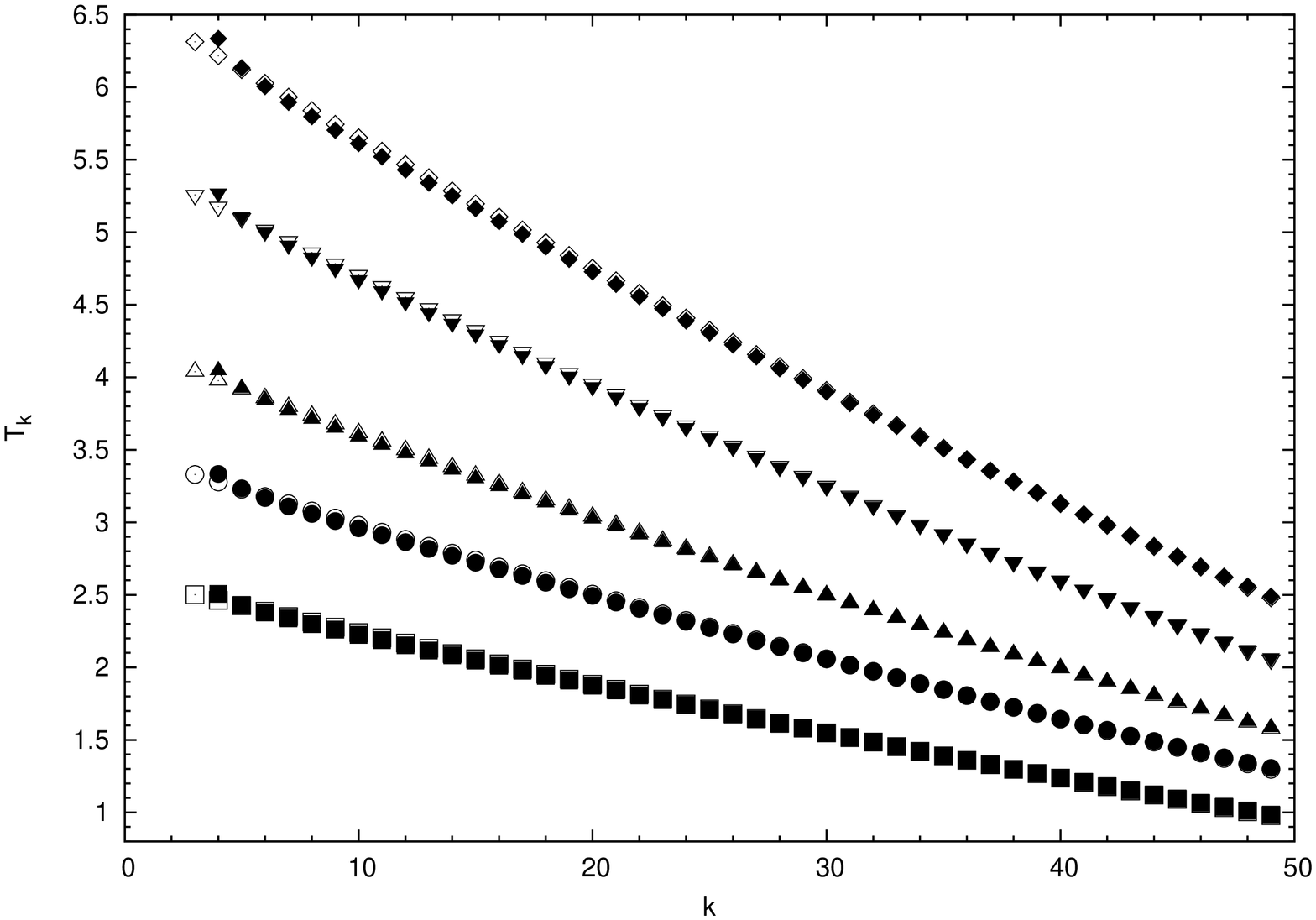}}}&
\rotatebox{-90}{\scalebox{.30}{\includegraphics{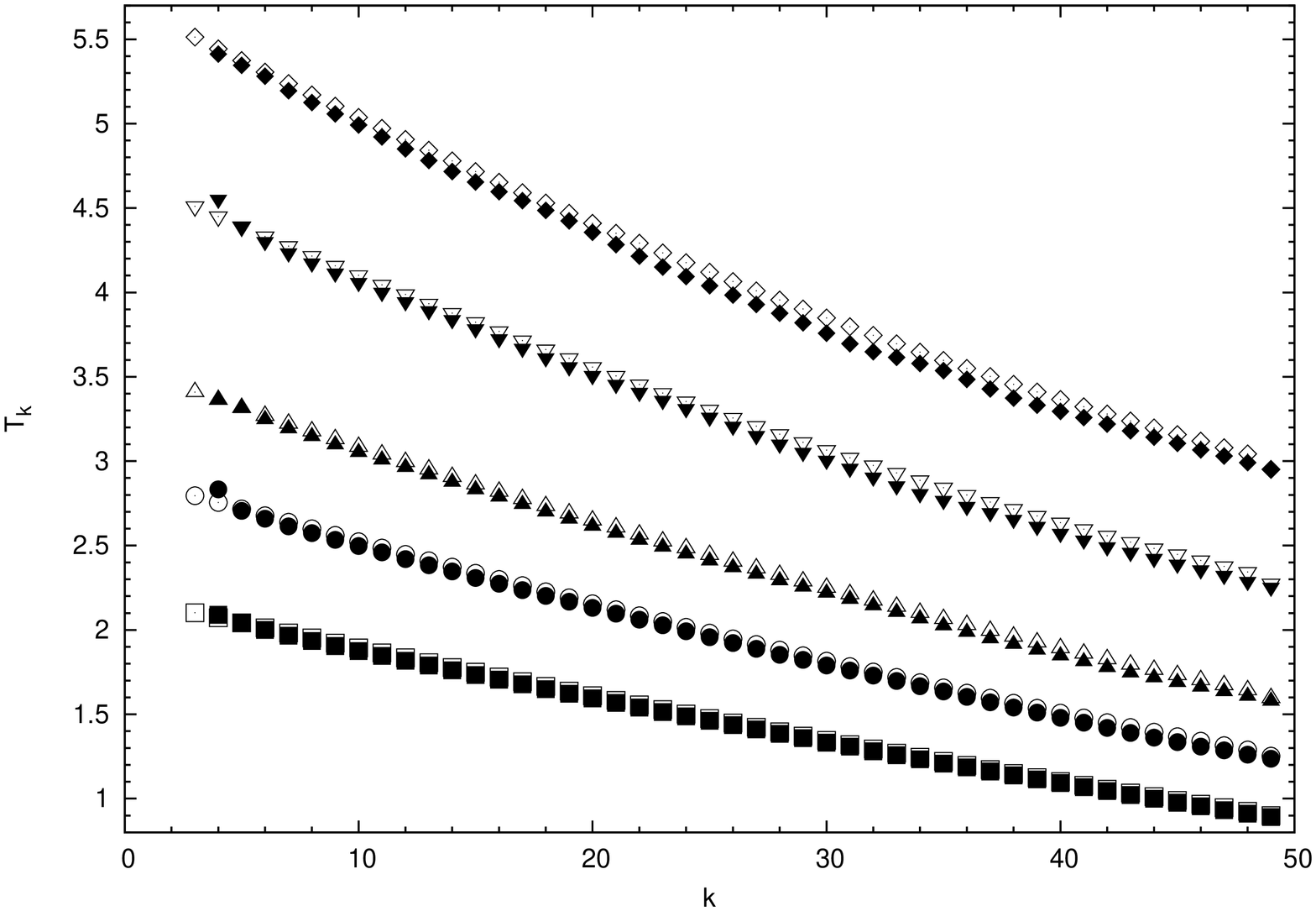}}}
\end{array}$
\caption{Residence time $T_{k}$ of the pulse on the $k$-th grain for $r=0.1$ (left panel) and 0.3 (right panel) for a chain with tapering parameter $S=0.01$. From bottom to top, the number of small grains between each pair of large ones increases as 1, 2, 3, 5 and 7. Numerical and analytical results are represented by the filled and open symbols respectively.}
 \label{fig:residencetime}
\end{figure*}

Although the binary collision approximation does not quantitatively capture the pulse amplitude, another important characteristic of the pulse is extremely well predicted. In Fig. \ref{fig:residencetime}, the residence time predicted by the binary collision approximation [Eq. (\ref{eq:residencetime})] and the one obtained from the direct integration of the equations of motion are shown to be approximately equal. Further, it is clear that the residence time increases with the number of small grains decorating the chain, that is, the pulse velocity decreases. It is also noticeable that the binary collision approximation is better for smaller decorating grains (the effective approach is ultimately based on the scale separation between large and small grains) and its quality decreases with the number of decorating grains due to error accumulation.
%%%%%%%%%%%%%%%%%%%%%%%%%%%%%%%%%%%%%%%%%%%%%%%%%%%%%%%%%%%%%%%%%%%%%%%%
%%%%%%%%%%%%%%%%%%%%%%%%%%%%%%%%%%%%%%%%%%%%%%%%%%%%%%%%%%%%%%%%%%%%%%%%%%%%

\section{Conclusions}
\label{sec:conclusion}
We have studied pulse propagation in granular chains decorated with a number of small particles between large ones. Using an effective description, the original chain was mapped into a non-decorated chain, for which we were able to apply the binary collision approximation to analytically calculate the residence time of the pulse on each effective large grain. We have successfully applied the theoretical predictions to forward decorated tapered chains. The use of other kinds of tapering is straightforward and should be equally successful as long as the ratio of the radii of the small grains to the smallest large grain is no larger than 0.4~\cite{Harbola-2}. Our description is also valid for monodisperse chains, for which the masses of the large grains are all equal and the time of residence is constant.

\section*{Acknowledgments}
A.R. acknowledges the financial support from CNPq and Capes/Nanobiotec. E.A.F.B. was supported by CNPq.

\end{document}